\documentclass[preprint,aps,floats,subeqn,showpacs]{revtex4}
\usepackage{epsfig}
\newcommand{\be}{\begin{equation}}
\newcommand{\ee}{\end{equation}}
\newcommand{\bea}{\begin{eqnarray}}
\newcommand{\eea}{\end{eqnarray}}
\newcommand{\bml}{\begin{mathletters}}
\newcommand{\eml}{\end{mathletters}}

\begin{document}

\tighten

%\preprint{IUB-TH-}
\draft

%\twocolumn[\hsize\textwidth\columnwidth\hsize\csname @twocolumnfalse\endcsname

%%%%%%%%%%%%%%%%%%%%%%%%%%%%%%%%%%%%%%%%%%%%%%%%%%%%%%%%%%%%%%%%%%%%%%%%%%

%\wideabs{    % Uncomment this line for two-column output

\title{The transition of 2-dimensional solitons to 1-dimensional ones on hexagonal
lattices}
\renewcommand{\thefootnote}{\fnsymbol{footnote}}
\author{Betti Hartmann\footnote{b.hartmann@iu-bremen.de}}
\affiliation{School of Engineering and Sciences, International University Bremen (IUB),
28725 Bremen, Germany}
\author{Wojtek J. Zakrzewski\footnote{w.j.zakrzewski@durham.ac.uk}}
\affiliation{Department of Mathematical Sciences, University of Durham,
Durham DH1 3LE, UK}

\date{\today}
\setlength{\footnotesep}{0.5\footnotesep}
\begin{abstract}
We study solitons arising in a system describing the
interaction of a two-dimensional discrete hexagonal lattice with an additional 
electron field (or, in general, an exciton field).
We assume that this interaction is electron-phonon-like. In our previous
paper \cite{hz}, we have studied the existence of two-dimensional solitons
 and have found that these solitons
 exist only if the electron-phonon coupling constant is sufficiently large. In this paper,
  we report the results of our investigation for small values 
  of this constant, close to its critical value for the existence of solitons.
We find that as the coupling decreases the soliton gets very broad and then becomes
  effectively one-dimensional.  
\end{abstract}
\maketitle

\section{Introduction}
Recently, a Fr\"ohlich Hamitonian was studied on 
a two-dimensional, discrete, quadratic \cite{bpz1,bpz3,bpz2}, respectively
hexagonal lattice \cite{hz}. 
It is well known that the interaction of
an excitation field such as an amide I- vibration in biopolymers
or an electron (in the case of the Fr\"ohlich Hamiltonian)
with a lattice whose distortion can be caused by this excitation
results in the creation of a localised state which, in what follows, we refer to as 
a soliton. Such a soliton was first 
introduced by Davydov \cite{davy} in the 1970s  to explain the
dispersion free energy transport in biopolymers (see also \cite{review} for
further details).
In \cite{bpz1,bpz3,hz}, the existence of such solitons 
was studied numerically and it was found that their properties
depended crucially on the magnitude of the electron-phonon coupling constant.
In particular, localised 2-dimensional structures exist only above a critical
value of this coupling. These numerical results have been confirmed by
a simple analytic argument: using a Gaussian in 2 dimensions, 
we can approximate 
very well this critical
value.

In this paper, we extend the results
for the soliton on a hexagonal lattice \cite{hz} 
and look at the properties of the solitons 
close to the critical coupling.  We find that close to this critical
coupling, three types of solutions exist: the ``narrow'' 2-dimensional solitons, discussed
in \cite{hz}, ``broad'' 2-dimensional solitons,
which extend over much bigger parts of the lattice than the former,
and one-dimensional solitons. Since our model is a toy-model
for solitons on nanotubes, these latter configuartions can be interpreted
as ring-like solitonic structures around the tube.
 
In section II we present our equations and in
section III we discuss the results of our numerical investigations.

\section{Basic Equations}
The equations describing an electron field
interacting with a deformable hexagonal lattice were derived in \cite{hz}.  These equations read:
\begin{eqnarray}
\label{eq1}
i \frac{\partial \psi_{i,j}}{\partial \tau}&=&(E_0+W_0)\psi_{i,j}-2
\left(\psi_{i+1,j+1}+\psi_{i-1,j}+\psi_{i+1,j-1}\right) \nonumber \\
&+&\psi_{i,j}\left[(U_{i+1,j+1}+U_{i+1,j-1}-2 U_{i-1,j})
+\sqrt{3}(V_{i+1,j+1}-V_{i+1,j-1})
\right] \ ,
\end{eqnarray}
for the electron field and
\begin{eqnarray}
\label{eq2}
\frac{d^2 U_{i,j}}{d\tau^2}&=&K_{x}\left(3U_{i,j}-U_{i+1,j+1}-U_{i-1,j}-U_{i+1,j-1}\right)\nonumber \\
&+& \frac{g}{2}\left(2\vert\psi_{i-1,j}\vert^2-\vert\psi_{i+1,j+1}\vert^2
-\vert\psi_{i+1,j-1}\vert^2\right)  \ , 
\end{eqnarray}
\begin{eqnarray}
\label{eq3}
\frac{d^2 V_{i,j}}{d\tau^2}&=&K_{x}\left(3V_{i,j}-V_{i+1,j+1}-V_{i-1,j}-V_{i+1,j-1}\right)\nonumber \\
&-& \frac{\sqrt{3}g}{2}\left(\vert\psi_{i+1,j+1}\vert^2 
-\vert\psi_{i+1,j-1}\vert^2\right)    \ 
\end{eqnarray}
for the displacement fields $U$ and $V$ in the $x$- and, respectively, 
$y$-direction.
 $i$ and $j$ index the points in $x$-, respectively $y$-directions.

For the corresponding Hamiltonian and the appropriate rescalings, we refer
the reader to \cite{hz}. Note that $K_x$ is the self-coupling constant
of the displacement fields and $g$ is the electron-phonon coupling.
$W_0$ is the phonon energy, while $E_0$ was chosen to be $0.142312$  
in our numerical investigations. We have normalised the electron-function
$\psi$ to one, i.e. $\sum\limits_{i,j=1}^{N_x,N_y} \psi_{i,j} \psi^*_{i,j} = 1$
 thus assuming the excitation to be one excess electron. Note that
$N_x$ and $N_y$ denote the number of lattice points in $x$-, respectively $y$-directions.

\section{Numerical results}
\begin{figure}[!htb]
\centering
\leavevmode\epsfxsize=12.0cm
\epsfbox{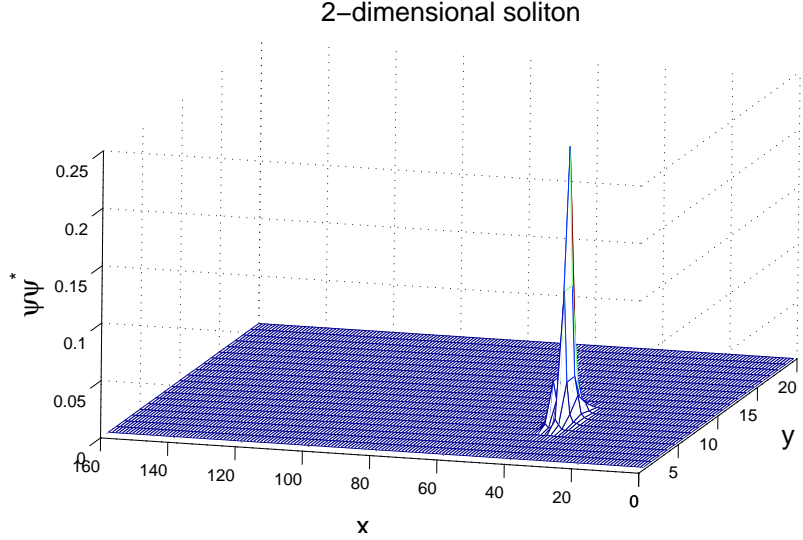}\\
\caption{\label{fignew1}  
A typical ``narrow'' 2-dimensional solitonic structure on a $N_x\times N_y=160\times 20$ lattice
for $g$ close
to the critical value $g_{cr}=2.296$.}
\end{figure}

 In this paper, we present new results on a model
describing the interaction of one excess electron with a 2-dimensional
hexagonal lattice. This model was studied first in \cite{hz}.
Before we discuss our new results, let us summarize the results of \cite{hz}.
In \cite{hz} it was found that 2-dimensional solitonic structures
which are created due to the interaction between
 the distortions of the hexagonal lattice (``phonons'')
and the electron field on this lattice (``excess electron''), exist only 
when the electron-phonon
coupling is large enough. This numerical result was also 
found in the case of the quadratic lattice \cite{bpz1,bpz2,bpz3} and thus
seems to be a generic feature of 2-dimensional lattices. For values of 
the coupling constant smaller than some critical value $g_{cr}$, the interaction
between the lattice and the electron (exciton) field becomes too small
for 2-dimensional solitonic structures to exist.
In this paper, we present new results on the behaviour of the system
close to this critical value $g_{cr}$. We have confirmed that
``narrow'', i.e. well pronounced 2-dimensional
solitons exist for $g > 2.296$. A typical solution of this type
is shown in Fig. \ref{fignew1} for $N_x=160$ and $N_y=20$.

\begin{figure}[!htb]
\centering
\leavevmode\epsfxsize=12.0cm
\epsfbox{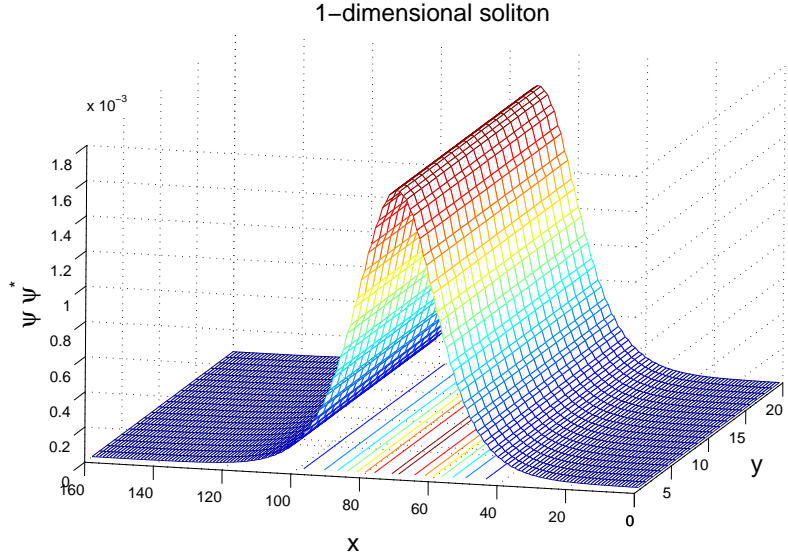}\\
\caption{\label{fignew2}  
A typical one-dimensional solitonic structure on a $N_x\times N_y=160\times 20$ lattice
for $g=2.295$.}
\end{figure}

\begin{figure}[!htb]
\centering
\leavevmode\epsfxsize=12.0cm
\epsfbox{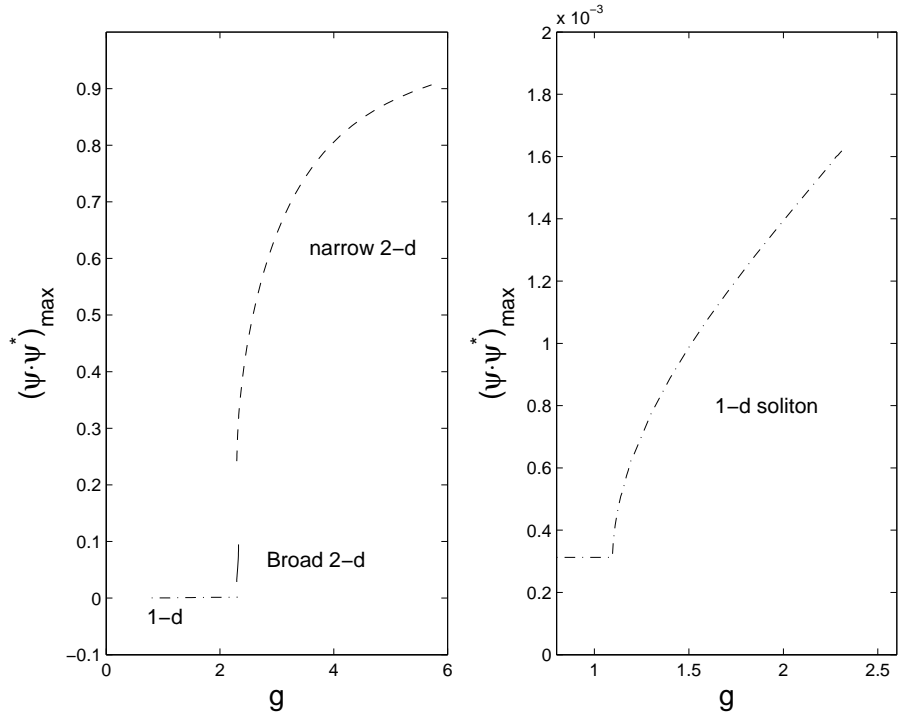}\\
\caption{\label{fig1}  
The $g$ dependence  of the height
$(\psi\psi^*)_{max}$ of the two-dimensional
and one-dimensional solitons.
Note that there exist ``narrow'' (dashed line) and 
``broad'' (solid line) two-dimensional solitons. The figure on the
right presents the zoom of the one-d soliton curve (dotted-dashed).}
\end{figure}

 Clearly the probability density $\psi\psi^*$ is localised only on a small
portion of the lattice.
For $g\ \epsilon\ [2.29475:2.32592]$, a broad two-dimensional soliton
appears (see Fig.\ref{fig1}), while decreasing the value of $g$ further leads
to a configuration in which the soliton becomes effectively one-dimensional.
A typical one-dimensional configuration is shown in Fig.\ref{fignew2}.

As is obvious from this figure, for small $g$, 
the function $\psi_{i,j}$ becomes effectively independent of
the $y$-direction and we see $N_y$ copies of the function $\psi_i$.
While for $g\rightarrow \infty$, 
 we  expect the solitonic structure to be effectively
fixed to one point on the lattice, thus $(\psi_{i,j} \psi^*_{i,j})_{max}=:
(\psi\psi^*)_{max}\rightarrow 1$,
the threshold of our one-dimensional soliton is characterised by
the value of $(\psi\psi^*)_{max}$, 
which can be obtained for
the last possible ``broad'' soliton divided by the number $N_y$. In our case
$N_y=20$ and the last value of the broad soliton
$(\psi\psi^*)_{max}\approx 2.84 \cdot 10^{-2}$, so
we would expect solitons with heights on the order
of $1.42\cdot 10^{-3}$.
Indeed, for $g < 2.29475$, we find solitons with heights of 
$\approx 1.62\cdot 10^{-3}$,
which confirms our interpretation.
We plot the $g$ dependence of the value 
of $(\psi\psi^*)_{max}$ in Fig.\ref{fig1}.
Clearly, we note the familiar structure of the curve, especially, we notice
that at $g\approx 1.0$, the value of $(\psi\psi^*)_{max}$ becomes constant
and is equal to $(\psi\psi^*)_{max}=3.125 \cdot 10^{-4}$. This is not surprising
since this is just the value where $\psi\psi^*=1/(N_x N_y)=1/(160\cdot 20)\approx 3.125 \cdot 10^{-4}$, i.e.
the field is completely delocalised, i.e. it is equally distributed over the 
whole grid. 

The fact that the one-soliton ceases to exist is, however, related to the number $N_y$.
In principle, one-d solitons should exist for all values of $ 2.29475 \ge g \ge 0$. Of course, this
value was derived for $\sum_{i,j}\psi_{i,j}\psi_{i,j}^*=1$.
To study the dependence on $N_y$, we have looked at the dependence of the 
critical value of $g$, $g_{cr,1d}(N_y)$ at which
the one-d soliton becomes equally distributed over the grid, i.e. 
$\psi_{i,j}\psi_{i,j}^*=1/(N_x N_y)$.
The values are listed in Table I.

\begin{table}[h]
\caption{Dependence of $g_{cr,1d}$ on $N_y$ for $N_x=160$}
\begin{ruledtabular}
\begin{tabular}{ccc}
$N_y$ &  $N_y^{-1}$ & $g_{cr,1d}(N_y)$  \\
20 & 0.050 & 1.00 \\
16 & 0.062 & 0.88\\
12 & 0.083 & 0.67\\
6 & 0.167 & 0.30 \\
\end{tabular}
\end{ruledtabular}
\end{table}

We notice that the value $g_{cr,1d}(N_y)$ decreases with increasing $N_y^{-1}$.
Since $N_y^{-1}$ represents the normalisation of the electron field
function in the $x$-direction, i.e. of the one-d soliton for one fixed $y$,
this means that the ``bigger'' the one-d soliton is the lower is the critical
value of the electron-phonon coupling. We can understand this behaviour
by observing that by scaling the $\psi$ field in our equations
(\ref{eq1})-(\ref{eq3})  such that $\psi_{i,j}\psi^*_{i,j}\rightarrow
N_y^{-1}(\psi_{i,j}\psi^*_{i,j})$ 
we obtain the equations for a system with a normalization of the
$\psi$ fields corresponding to the one-d solitons.
We notice that this rescaling is nothing else but the rescaling of the
coupling constant $g\rightarrow N_y^{-1} g$. 
Following this argument, the product $N_y^{-1} g_{cr,1d}$ should thus be
constant. Looking at Table I, we indeed find that $N_y^{-1} g_{cr,1d}\approx 0.05$.

\begin{figure}[!htb]
\centering
\leavevmode\epsfxsize=12.0cm
\epsfbox{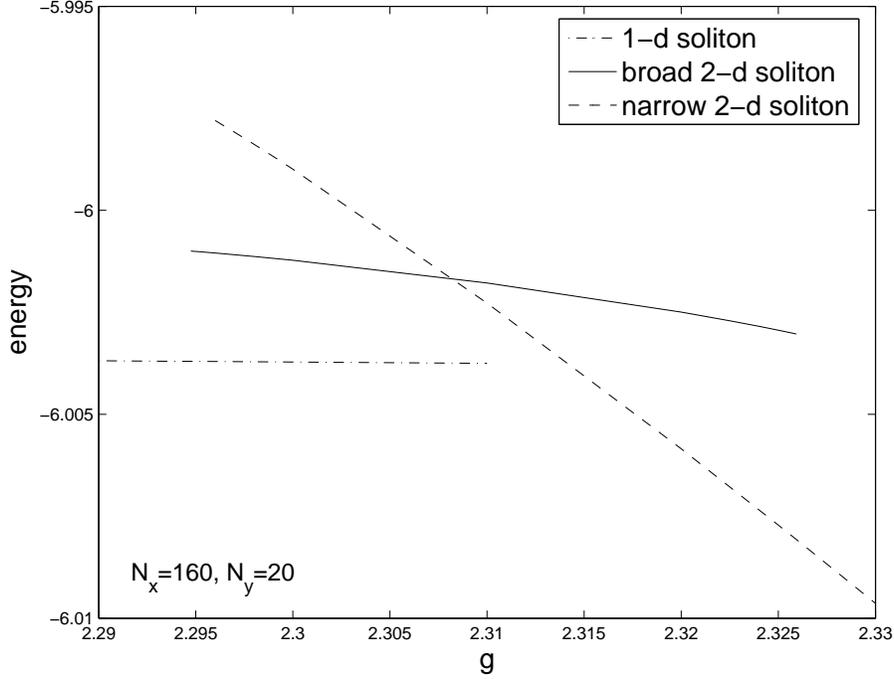}\\
\caption{\label{fig2} The energy of the two-dimensional narrow, respectively, broad solitons
in comparsion with the energy of the one-d soliton.
Note that the energy of the configuration with equally distributed 
$\psi_{i,j}\psi_{i,j}^*=1/(160\cdot 20)$ would have energy $\approx -5.8576$. 
}
\end{figure}

We have also plotted the corresponding energies of the solutions.
These are shown in Fig.\ref{fig2}. Note that the configurations with
equally distributed $\psi$ would have an energy $\approx -5.8576$.
The energies of the soliton solutions are clearly below this value.
The energy of a narrow soliton is the lowest energy configuration 
as long as the coupling constant is above the value at which one-d solitons can exist.
Then, these one-d solitons have the lowest energy. Note that the broad soliton
never becomes the lowest energy configuration and so appears to be unstable.

\section{Conclusions}

In this short note we have reported results of our studies of an electron phonon system on a hexagonal
lattice which was longer in one direction than the other (i.e. with a larger number of
lattice points in the $x$-direction - $N_x$ than in the $y$-direction i.e. $N_y$). We have looked at the values
of the electron-phonon coupling constant at which the system possesses soliton-like
solutions. We have found that ``genuine'' two-dimensional solitons, like those studied
in \cite{hz}, exist for the coupling constant larger than some critical value.
For the values of the constant close to this critical value the system possess another
solution - which corresponds to a ``broad'' soliton. This new solution appears to be unstable.
When the coupling constant is below its critical value the two-dimensional
soliton is so ``broad'' that it can be interpreted as an effective
one-dimensional structure. As the value of the coupling decreases 
further the ``one-dimensional'' soliton becomes broader and, below some new critical value,
it spreads over the whole lattice leading to complete delocalisation.

The value at which this happens depends on the size of the lattice. However, this is not surprising
as for an effective one-dimensional soliton the normalisation of the electron field
plays an important role (the soliton is normalised on the two-dimensional lattice and
so the field  of the effective one-dimensional soliton is normalised to $1\over N_y$, 
where $N_y$ describes the number of points in the ``shorter'' direction). A simple scaling argument
then explains the observed $N_y$ dependence of the critical value of the electron-phonon
coupling constant for the existence of one-dimensional solitons.

 Though our investigations were performed only for the 2-dimensional
hexagonal lattice, we would expect that similar features appear in the
case of other lattice structures, 
i.e. lattices in physical three dimensional spaces, the quadratic lattices of \cite{bpz1,bpz2,bpz3} etc.

\acknowledgements{We would like to thank L. Brizhik and A. Eremko for helpful
comments and suggestions.}

\end{document}